\documentclass[12pt,onecolumn]{revtex4}

\usepackage{graphicx}
\usepackage{amsmath}
\usepackage{amssymb}
\usepackage{mathrsfs}
\usepackage{color}
\usepackage{multirow}
\usepackage{epstopdf}
\usepackage{textcomp}

\usepackage{ulem}

\usepackage{}

\begin{document}

\title{All-optical spatio-temporal metrology for isolated attosecond pulses}

\author{Lixin He,$^{1,2}$ Jianchang Hu,$^{1}$  Siqi Sun,$^{1}$ Yanqing He,$^{1}$ Yu Deng,$^{1}$ Pengfei Lan,$^{1,2,}$\footnote{pengfeilan@hust.edu.cn} and Peixiang Lu$^{1,2,3}$}

\affiliation{%
$^1$Wuhan National Laboratory for Optoelectronics and School of Physics,
Huazhong University of Science and Technology, Wuhan 430074,
China\\
$^2$ Optical Valley Laboratory, Hubei 430074, China\\
$^3$CAS Center for Excellence in Ultra-intense Laser Science, Shanghai 201800, China\\
}%

\begin{abstract}
Characterizing an isolated attosecond pulse (IAP) is essential for its potential applications. A complete characterization of an IAP ultimately requires the determination of its electric field in both time and space domains. However, previous methods, like the widely-used RABBITT and attosecond streaking, only measure the temporal profile of the attosecond pulse.
Here we demonstrate an all-optical method for the measurement of the space-time properties of an IAP. By introducing a non-collinear perturbing pulse to the driving field, the process of IAP generation is modified both spatially and temporally, manifesting as a spatial and a frequency modulation in the harmonic spectrum. By using a FROG-like retrieval method, the spatio-spectral phases of the harmonic spectrum are faithfully extracted from the induced spatio-spectral modulations, which allows a thoroughgoing characterization of the IAP in both time and space.
With this method, the spatio-temporal structures of the IAP generated in a two-color driving field in both the near- and far-field are fully reconstructed, from which a weak spatio-temporal coupling in the IAP generation is revealed. Our approach overcomes the limitation in the temporal measurement in conventional $in$ $situ$ scheme, providing a reliable and holistic metrology for IAP characterization.
\end{abstract}                         %

\maketitle

\section{Introduction}

The advent of attosecond extreme ultraviolet (XUV)/soft X-ray pulses via high-order harmonic generation (HHG) is a milestone in strong-field physics and attoscience \cite{atto1,atto2,atto3,bie,atto4,chang,worner,xue}, which has opened up new avenues for accessing ultrafast electron dynamics in atoms \cite{dy1,dy2,dy3,dy4}, molecules \cite{dy5,dy6,dy7}, and condensed matter \cite{dy8} on its natural time scale.  HHG is a highly nonlinear process during the laser-matter interaction \cite{hhg1,hhg2,hhg3}, accompanying with complicated macroscopic effects in the propagation \cite{propa1,propa2,propa3}.
Isolated attosecond pulses (IAPs) produced by HHG generally have complex spatio-temporal structures, which encode both the \aa ngstrom-sized spatial features and attosecond scale
temporal features of the response of the matters to the laser field.
A complete characterization of the IAP in both time and space is critical
not only for the development of new attosecond light sources, but also for its applications in attosecond pump-probe
experiments, as well as for unraveling
the physics underlying the laser-matter interaction \cite{cha1,cha2,cha3}.

The complete characterization of an ultrashort IAP actually requires the
determination of its spatio-temporal electric field $E(x,y,t)$, or its spatio-spectral counterpart $\tilde{E}$ $(x,y,\omega)$. To date, 
attosecond streaking technique \cite{str0,str00} has been usually used to retrieve the temporal profile of an IAP from the streaked photoelectron spectrogram of atoms with the FROG-CRAB (Frequency-resolved optical
gating for complete reconstruction of attosecond bursts) algorithm \cite{str1,str2,str3,str4}. Several other methods, such as the Phase retrieval by omega
oscillation filtering (PROOF) \cite{P1}, Volkov transform generalized projection algorithm (VTGPA) \cite{P2}, the genetic
algorithms \cite{P3,P4,P5}, and the neural
networks \cite{P6,P7}, have also been developed to overcome the  theoretical and experimental limitations in the attosecond streaking technique, e.g., the assumption of central momentum approximation (CMA) and the high statistic noise in the streaking trace due to the low photon flux of the attosecond pulse. However, all these temporal characterization approaches rely on the conversion of the
attosecond pulse into electron wave-packets through photoionization of atoms, which have automatically averaged the spatial structure of the IAP in the measurement. Considering the possible significant space-time coupling in the highly nonlinear generation process of IAP \cite{st1,st2}, the temporal measurement averaged over the space will be inadequate in many cases.  Although some approaches, such as, the point-diffraction
interferometry (PDI) \cite{spa1}, the spectral wavefront optical
reconstruction by diffraction (SWORD) \cite{spa2}, and the lateral shearing
interferometry (LSI) \cite{spa3}, have been demonstrated for the spatial characterization of high-order harmonics, none of them is compatible with the temporal measurement.

On the other hand,  an $in$ $situ$ method based on a weak perturbation of the harmonic generation process by a second-harmonic field that co-propagates with the fundamental beam was introduced \cite{DD}. 
In this scheme, the temporal information of attosecond pulse was decoded from the two-color delay-dependent modulations of the generated even-harmonics. Later on, Kim \textit{et al.} reformed this $in$ $situ$ method with the weak perturbing beam propagating non-collinearly with the driving beam \cite{kim}. This configuration built a spatial gating as the two-color delay varies, allowing a synchronous measurement of the spatial and temporal characteristics of the attosecond pulse. However, in this $in$ $situ$ scheme, a linear spectral group delay dispersions (GDD) of the emitted
harmonics were assumed and derived from the spatial modulation of the far-field harmonic signals. More complex spectral phase information, such as high-order dispersion in the IAP, is critical for the temporal measurement but is unavailable with this scheme.




In this work, we demonstrate an improved $in$ $situ$ approach for accurate measurement of the spatio-temporal structures of an IAP. We find that in addition to the spatial modulation, the utilization of the weak perturbing beam can also lead to a frequency (temporal) modulation in the harmonic generation as its time delay varies. With a method similar to the phase retrieval of femtosecond laser with the frequency-resolved
optical gating (FROG) technique \cite{FROG}, the spatial and spectral harmonic phases are successfully extracted from the spatial and frequency modulations of the generated harmonics, respectively. This allows us to build the spatio-spectral connection of the harmonic phase and then achieve a complete characterization of IAPs in both time and space. Our method obviates the linear GDD approximation in conventional $in$ $situ$ measurement \cite{kim}, allowing an exact characterization of the IAP. With this method, we have demonstrated a full spatio-temporal reconstruction of the IAP generated by a two-color driving field in experiment. The spatio-temporal coupling effect during the IAP generation is revealed from the reconstructions.

\begin{figure*}[hbt]
	\centerline{
		\includegraphics[width=15cm]{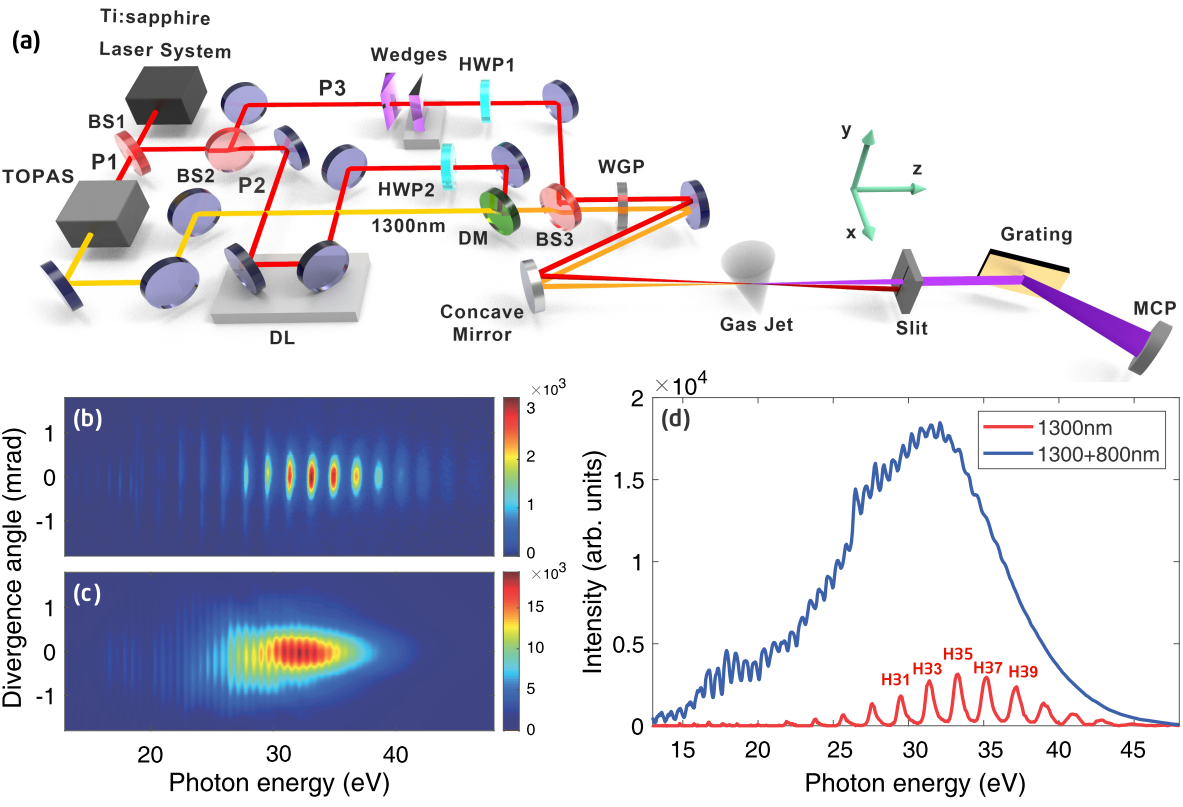}} \caption{\label{fig1} (a) Schematic diagram of the experimental set-up for the spatio-temporal measurement of the IAP generated in a tow-color driving field. BS, beam splitter; DM, dichroic mirror; DL, delay line; HWP, half-wave plate; WGP, wire grid polarizer. (b) Spatially-resolved harmonic spectrum driven by the 1300 nm fundamental field alone. (c) Same as (b), but for the two-color driving field synthesized by the 1300 nm fundamental and 800 nm assistant laser fields. (d) On-axis harmonic spectra driven by the one-color (red line) and two-color (blue line) laser field, respectively.}
\end{figure*}

\section{Methods}

\subsection{Principle of the spatio-temporal characterization of an IAP}

For the spatio-temporal characterization of an IAP, we  imitate the spatially encoded $in$ $situ$ method in \cite{kim} by introducing a weak perturbing beam to modify the HHG process both spatially and temporally. In space, the perturbing beam could alter the wavefront of HHG in the near-field, which, in turn, will modify the far-field
spatial distribution of the harmonic spectrum. By scanning the 
time delay between the perturbing and driving lasers, spatial modulations in the far-field distributions of the harmonics can been observed. In our experiment [see Fig. 1(a)], the lasers have a transverse spatial distribution in the ($x$,$y$) plane normal to the propagation direction $z$. Only the vertical component of the far-field pattern of the harmonic spectrum is measured due to the use of an entrance slit along the $y$ direction. Based on the strong field approximation (SFA) theory \cite{hhg2,DD,kim}, the spatial modulation induced by the perturbation in our experiment can be approximated as
\begin{eqnarray}
|E_{far}(\omega,\theta,\tau)|^2\approx|\int E_{near}(\omega,y,\tau)G(y-c\tau/\vartheta)e^{ik_{\omega}\theta y}dy|^2.
\end{eqnarray}
where $E_{far}$ and $E_{near}$ are spatially-resolved complex amplitude of the harmonic with the frequency of $\omega$ in the far- and near-field, respectively. The modulus squares of $E_{far}$ and $E_{near}$ represent the distributions of the harmonic in space. $y$ is the spatial coordinate in the near-field, $\theta$ is the divergence of the harmonic in the far-field, $\tau$ and $\vartheta$ are the time delay and crossing angle between the perturbing and driving lasers, respectively.
$k$$_{\omega}$ is the wavenumber of the harmonic, $c$ is the speed of the light. 
$G(y-c\tau/\vartheta)$ is a spatial gate induced by the perturbing pulse, which represents the modifications in both the amplitude and phase of $E_{near}$. Note that, in Eq. (1), only the short quantum trajectory is considered due to its better phase matching in our experiment.
In this case, reconstructing the spatial structure of the unperturbed harmonics ($E_{near}$) is equivalent to solving
the retrieval problem of femtosecond laser pulses with the FROG technique \cite{FROG}. 
By using the standard FROG algorithm, e.g., the widely-used  principal component
generalized projection algorithm (PCGPA) \cite{PCGPA},
both the unperturbed harmonics ($E_{near}$) and the spatial gate $G(y-c\tau/\vartheta)$ can be uniquely extracted from the delay-dependent far-field patterns of the harmonics.

The second step is to determine the temporal properties of the IAP. In this regard, it needs to know the relative spectral phase of the whole spectrum. 
In \cite{kim}, this relative spectral phase has been defined as the linear group delay in HHG, and was directly acquired from the delay-dependent spatial modulations of harmonics with different photon energies. This manner ignores more complex phase information in the IAP, which will limit the accuracy of the  temporal measurement. Here we demonstrate an improved method for accurate temporal measurement.
We find that, in addition to the spatial modulation, the introduced weak perturbing pulse can also lead to a modulation in the HHG process in the time domain, i.e.,
\begin{eqnarray}
d(t,\tau)=d_0(t)g(t,\tau)+c.c.
\end{eqnarray}
where $d_0(t)$ denotes the dipole moment without the perturbation, $g(t,\tau)$ is the modulations induced by the perturbing pulse. In previous works \cite{DD,cao}, it was usually assumed that the weak perturbing pulse will not
affect the ionization process, but mainly perturbs the free electron’s action in the laser field. Then the induced modulations can be approximated as $g(t,\tau)$=$e^{-i\sigma(t,\tau)}$, where $\sigma(t,\tau)$ is the additional phase induced by the perturbing pulse. In this work, the influence of the perturbing pulse on both the dipole amplitude and phase have been taken into account. That means $g(t,\tau)$ in our reconstruction is a complex-valued quantity.
The harmonic spectrum then can be obtained by 
\begin{eqnarray}
S(\omega,\tau)=\omega^4|\int d_0(t)g(t,\tau) e^{i\omega t}dt |^2.
\end{eqnarray}
Eq. (3) indicates that the weak perturbing pulse introduces a temporal phase gate in the harmonic generation, which will lead to a frequency modulation in the harmonic spectrum. Reformulating Eq. (3), we can get
\begin{eqnarray}
\frac{S(\omega,\tau)}{\omega^4} = |\int d(t,\tau) e^{i\omega t}dt |^2 = |\int d_0(t)g(t,\tau) e^{i\omega t}dt |^2.
\end{eqnarray}
Eq. (4) is also equivalent to the equation used in FROG technique. It allows us to retrieve the unperturbed complex dipole moment $d_0(t)$ and then the spectral phase from the frequency-modulated harmonic spectra by using PCGPA. With both the spatial and spectral phases reconstructed, we can then achieve a full characterization of the IAP in both time and space.

\begin{figure*}[hbt]
	\centerline{
		\includegraphics[width=15cm]{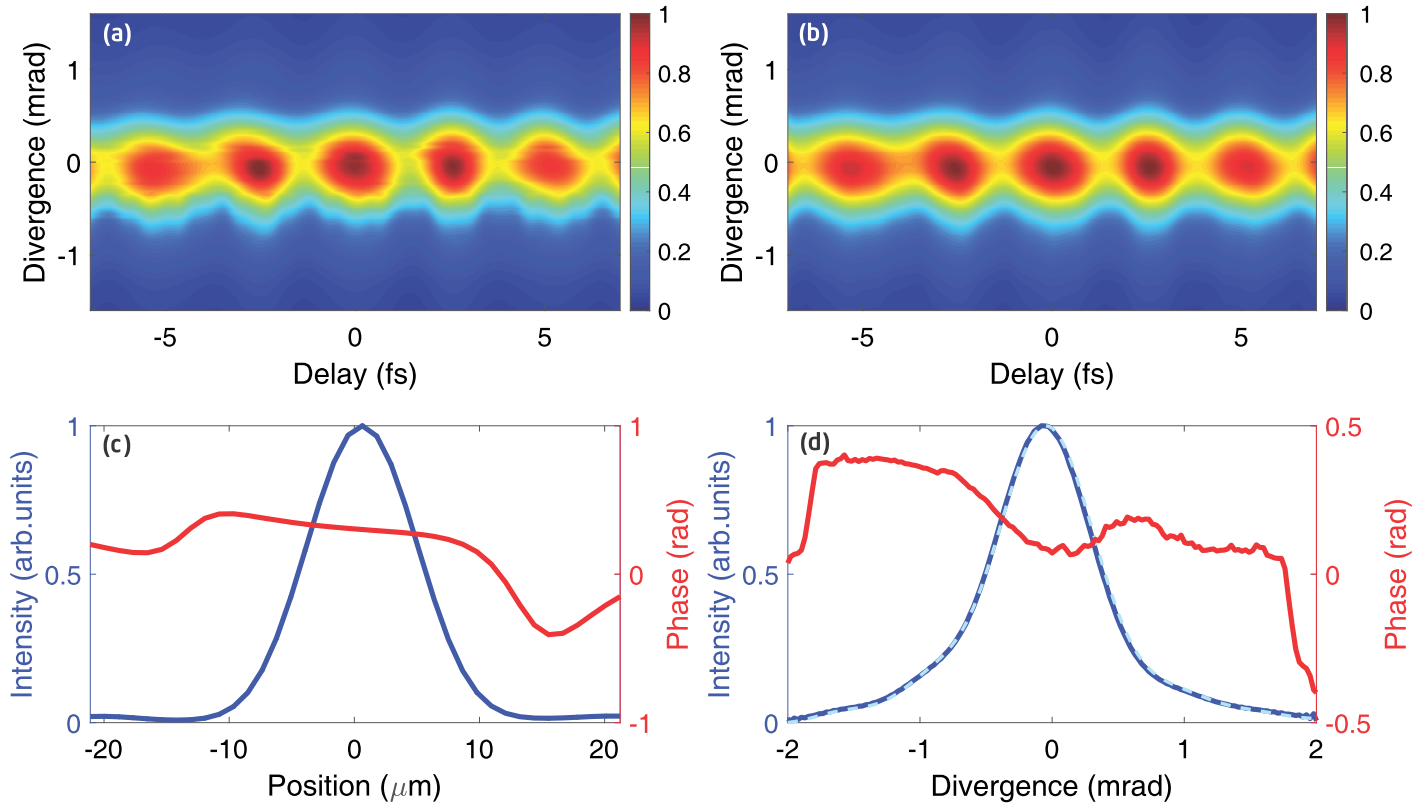}}
	\caption{\label{fig2} (a) Far-field spatial distribution of H35 as a function of time delay of the perturbing laser. (b) Same as (a), but for the PCGPA reconstruction. (c) Reconstructed near-field distribution of  the intensity (blue line) and phase (red line) of H35. (d) Reproduced far-field intensity (blue line) and phase (red line) of H35. The cyan dashed line plots the measured far-field intensity of the unperturbed H35 for comparison.}
\end{figure*}

\subsection{Experimental method}
In our work, to obtain an IAP, we adopted a two-color driving field, which is synthesized by a 1300 nm fundamental field and an 800 nm assistant field, to generate high-order harmonics. A schematic diagram of the experimental setup is shown in Fig. \ref{fig1}(a).  To be specific, the output of a commercial Ti: sapphire laser system (Astrella-USP-1K, Coherent, Inc.), which delivers 33 fs, 800 nm pulses at a repetition rate of 1 kHz, with the maximum pulse energy of 7 mJ, was divided into three beams by two 30\% reflection beam splitters (BS1 and BS2). The transmitted beam P1 after BS1 with the energy of 4.9 mJ was employed to pump an optical parametric amplifier (TOPAS-Prime-Plus, Coherent) to generated a 60 fs, 1300 nm fundamental driving laser. The transmitted 800 nm laser P2 after BS2 was recombined with the 1300 nm laser by a dichroic mirror (DM) as an assistant field to synthesize the two-color driving field. A motorized delay line was installed in the arm of P2 to control the relative phase of the two-color field.

A weak 800 nm laser P3 reflected by BS2 was introduced as a perturbing beam to modify the HHG process. The two-color driving field and the perturbing beam were recombined by BS3 and focused non-collinearly (the crossing angle is about 15 mrad) onto a gas jet to generate high-order harmonics by a f=20 cm concave mirror. In our experiment, we chose argon atom as the prototype to demonstrate our scheme. The background gas pressure is about 20 Torr. The gas jet is placed at 1mm after the laser focus to ensure the phase matching of the short quantum trajectory. A wire grid polarizer (WGP) was inserted after BS3 to ensure the same polarization direction of all these pulses. In combination with the WGP, two half-wave plates (HWPs) were installed in the arms of P2 and P3 to adjust their intensities. A pair of wedges were installed in the arm of P3 to adjust the time delay between the perturbing pulse and the two-color field.

The generated high-order harmonics were detected by a homemade flat-field soft x-ray spectrometer, which consists of a 0.2-mm-wide, 15-mm-height entrance slit, a flat-field
grating (1200 grooves mm$^{-1}$), and a microchannel plate (MCP) backed with a phosphor screen. A charge-coupled device (CCD) camera is used to record the spectrally resolved images.

\begin{figure}[hbt]
	\centerline{
		\includegraphics[width=10cm]{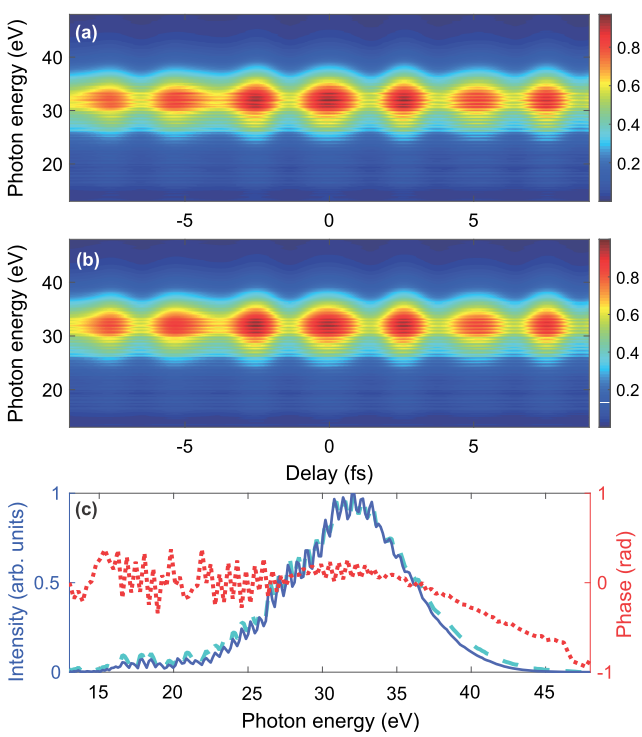}}
	\caption{\label{fig3} (a) On-axis harmonic spectra measured in the two-color driving field as a function of time delay of the  perturbing laser. (b) Same as (a), but for the PCGPA reconstruction. (c) Reproduced intensity (blue solid line) and phase (red dotted line) of the unperturbed on-axis harmonic spectrum in the far-field. The cyan dashed line plots the far-field on-axis harmonic intensity measured without the perturbation for comparison.}
\end{figure}

\section{Results and discussion}

Figure \ref{fig1}(b) plots the measured far-field pattern of the harmonic spectrum driven by the 1300 nm fundamental laser alone. One can see clear discrete odd harmonics with the photon energy up to 45 eV [47th harmonic (H47)] in the spectrum [see red line in Fig. \ref{fig1}(d)]. These discrete harmonics are associated with the bursts of attosecond pulses within each half optical cycle, i.e., an attosecond pulse train in time domain. While in the two-color driving scheme, by optimizing the intensity and relative phase of the 800 nm assistant field with respect to the 1300 nm fundamental field, a smooth supercontinuum in the region of 33-45 eV is obtained [see Figs. \ref{fig1}(c)-(d)], which supports the generation of an IAP in the experiment. Small modulation that appears in the range of 13-33 eV indicates the existence of satellite attosecond pulses beside the main pulse as in \cite{tk}. Moreover, as shown in Fig. \ref{fig1}(d), the HHG yield in the two-color field (blue line) is visibly enhanced (by $\sim$ 5 times) compared to that in the one-color field (red line) due to the increase of the ionization rate induced by the 800 nm assistant field.


Figure \ref{fig2}(a) shows the far-field spatial distributions of H35 measured in the two-color laser field as a function of the time delay between the perturbing and driving laser pulses. One can see clear modulations in the far-field distribution of the harmonic as the time delay varies. According to Eq. (1) , the unperturbed complex harmonic amplitude in the near-field as well as the spatial gate induced by the perturbing beam can be uniquely extracted from the FROG-like trace in Fig. \ref{fig2}(a) by using the PCGPA method.

Figure \ref{fig2}(b) displays the PCGPA reproduction of the far-field pattern of H35 as a function of time delay of the perturbing pulse. The result agrees well with the experimental measurement shown in Fig. \ref{fig2}(a). Figure \ref{fig2}(c) plots the reconstructed distributions of the intensity (blue line) and phase (red line) of the unperturbed H35 in the near-field. With the reconstructed near-field result, we can further obtain the far-field distribution of the unperturbed harmonic in terms of 
\begin{eqnarray}
E_{far}(\omega,\theta)=\int E_{near}(\omega,y,\tau)e^{ik_{\omega}\theta y}dy.
\end{eqnarray}
The reconstructed far-field results are plotted in Fig. \ref{fig2}(d). For comparison, we have also performed HHG experiment without the perturbation. The measured far-field distribution of the intensity of H35 is plotted as the cyan dashed line in Fig. \ref{fig2}(d), which shows good agreement with the reconstruction (blue solid line), proving the validity of our reconstruction.  Repeating the above procedure for each photon energy, we can then achieve the spatial measurement of  both the amplitude and phase of the whole harmonic spectrum in both the near- and far-field. 

Secondly, we demonstrate the temporal measurement of the generated IAP. To this end, it requires to determine the relative spectral phase of the whole harmonic spectrum. Our method relies on the frequency (temporal) modulation induced by the perturbing laser in the harmonic generation. As shown in Fig. \ref{fig3}(a), by scanning the time delay between the perturbing and driving laser fields, one can see clear modulations in the 2-D spectrogram of the far-field on-axis harmonic spectrum. Similar result is also  found for the harmonic spectra at other spatial positions (not presented here). According to Eq. (4), we can retrieve the unperturbed complex harmonc dipole moment $d_0(t)$ and then the spectral phase of the harmonic spectrum from the measured 2-D spectrogram trace by using the PCGPA method.

\begin{figure*}[hbt]
	\centerline{
		\includegraphics[width=14cm]{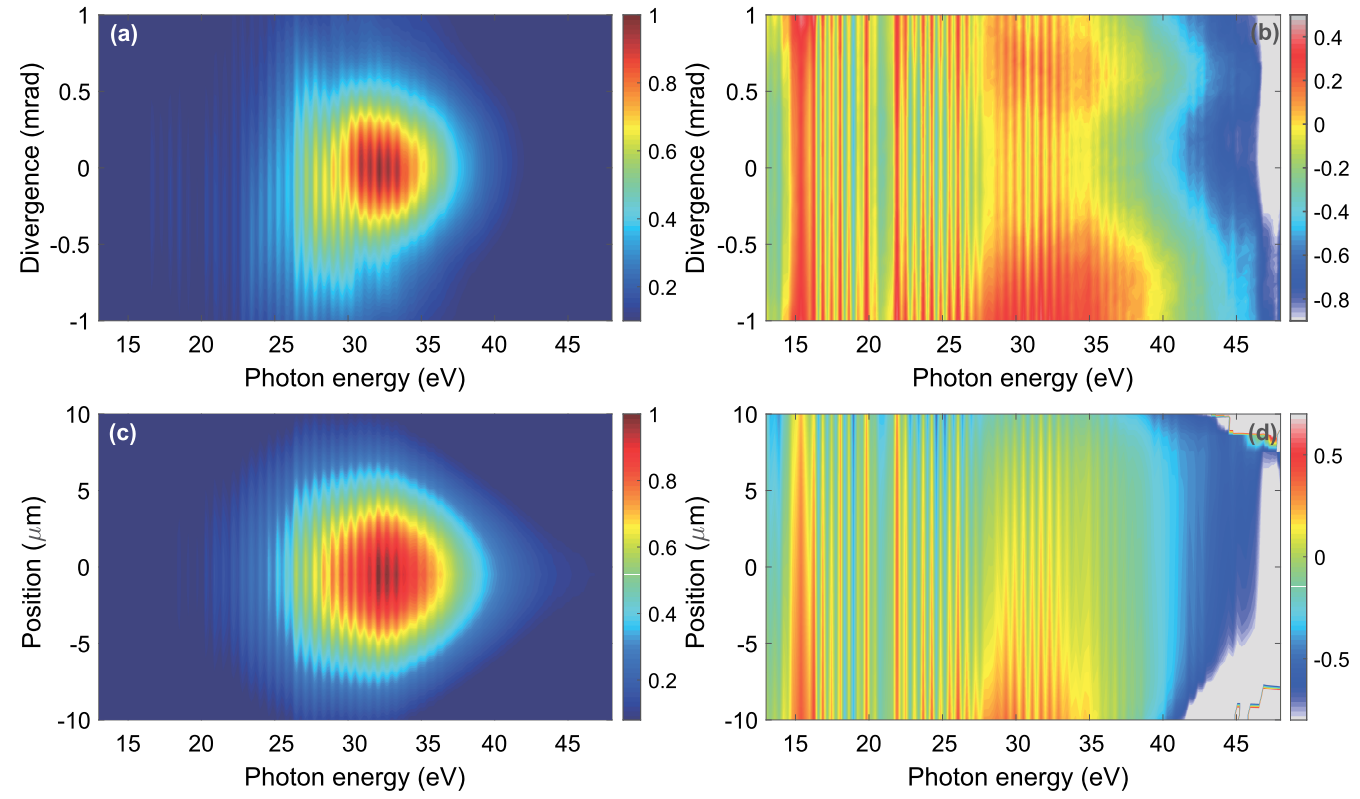}}
	\caption{\label{fig4} (a)-(b) Spatially-resolved far-field harmonic intensity (a) and phase (b) reconstructed from the measurement. (c)-(d) Same as (a)-(b), but for the near-field result.}
\end{figure*}

In Fig. \ref{fig3}(b), we plot the PCGPA reproduction of the spectrogram trace of the far-field on-axis harmonic spectrum. As shown, the reconstruction agrees well with the measurement in Fig. \ref{fig3}(a). Figure \ref{fig3}(c) displays the reconstructed unperturbed spectral intensity (blue solid line) and phase (red dotted line) of the on-axis harmonic spectrum in the far field. The reconstructed spectral intensity is also in good agreement with the unperturbed measurement (cyan dashed line). 
In the supercontinuum, with the photon energy varying from 33 eV to 47 eV, the harmonic phase has changed by 0.75 rad, corresponding to a 320 as difference in the emission times of these harmonics.
Besides, one can see that the spectral phase is irregular 
in the energy range from 13 eV to 33 eV, which deviates far from the linear group delay defined in \cite{kim}, indicating a complex dispersion in this energy range. This complex dispersion can not be measured with the method in \cite{kim}.
By connecting the retrieved spectral phase and the spatial reconstructions above, the intensity and phase of the far-field harmonics then can be fully determined in the whole space. The near-field result can also be obtained by performing inverse Fourier transform of the far-field result. 
Note that the relative spectral phase in principle can be retrieved from the harmonic spectrum measured at each spatial position in the far-field. Owing to the experimental errors in the measurement, the phase connection built at different spatial positions may be discrepant. In this work, we have retrieved the spectral phase from the far-field harmonic spectra in the range from -0.5 to 0.5 mrad, and then optimized the spatio-spectral phase connection from the retrievals by minimizing the difference between the measurement and the reconstruction with a  surrogate-based optimization algorithm \cite{su}. Figure \ref{fig4} shows the reconstructed distributions of the harmonic intensity [Fig. \ref{fig4}(a) and (c)] and phase [Fig. \ref{fig4}(b) and (d)] in both the far- and near-field.

\begin{figure*}[hbt]
	\centerline{
		\includegraphics[width=14cm]{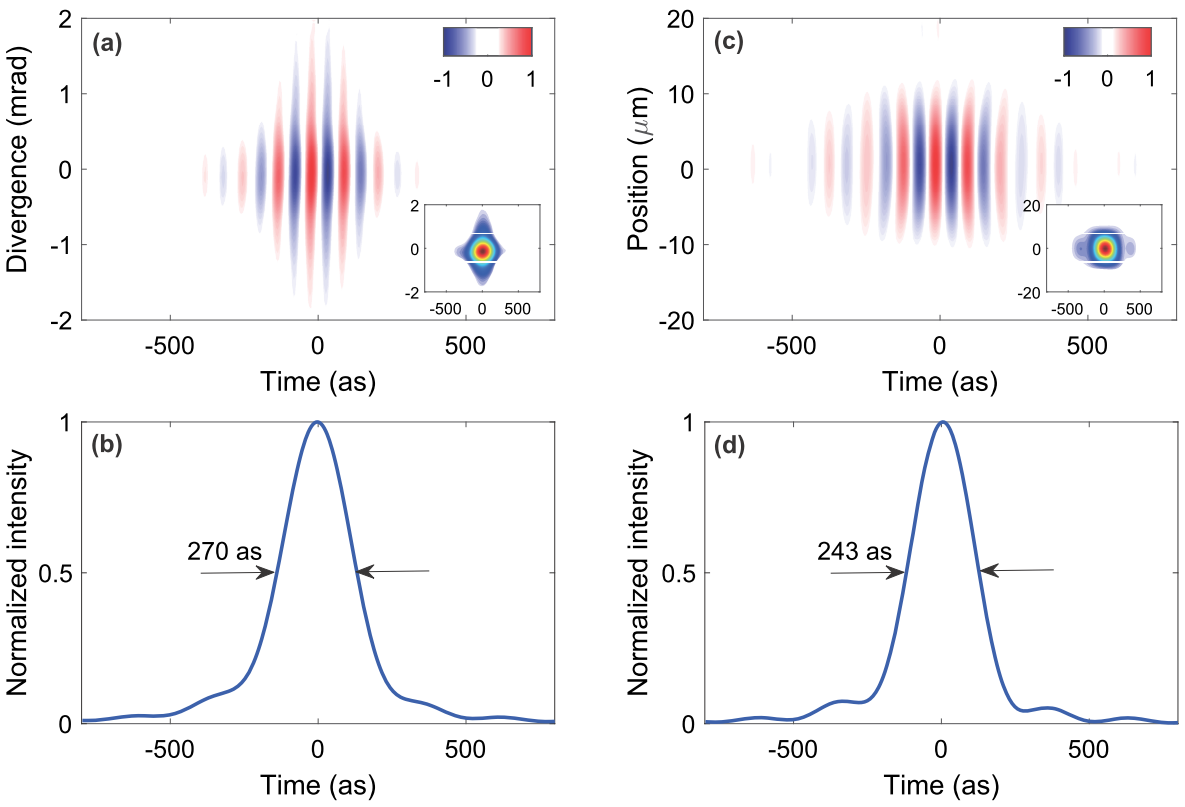}}
	\caption{\label{fig5} (a) Spatio-temporal profiles of the electric field of the attosecond pulse in the far-field obtained by synthesizing harmonics in the energy range from 33 eV to 47 eV. The inset shows the corresponding spatio-temporal distribution of the intensity of the attosecond pulse. (b) Temporal profile of the on-axis attosecond pulse in the far field. (c)-(d) Same as (a)-(b), but for the near-field result.}
\end{figure*}

With the reconstructed harmonic amplitudes and phases in Fig. \ref{fig4}, we can finally achieve a spatio-temporal measurement of the IAP by superposing the harmonics in the supercontinuum (33-47 eV) coherently. Figure \ref{fig5}(a) shows the electric field of the far-field attosecond pulse as a function of the time and position. The inset plots the spatio-temporal profile of the pulse intensity. The pulse duration of the on-axis emission is measured to be 270 as [ Fig. \ref{fig5}(b)]. The results in the near-field are also shown in Figs. \ref{fig5}(c)-(d). The duration of the on-axis attosecond pulse in the near-field is about 243 as, which is slightly shorter than the far-field result. 


For a quantitative characterization of the spatio-temporal structure of the generated IAP,  we have further calculated the normalized spatial chirp $\rho_{y \omega}$ of the IAP in both the near- and far-field in terms of \cite{chirp1,chirp2}
\begin{eqnarray}
\rho_{y \omega}=\frac{\iint{y \omega I(y, \omega)  dy d\omega}}{\sqrt{\iint{y^2 I(y, \omega)  dy d\omega}
		\iint{\omega^2 I(y, \omega)  dy d\omega}}},
\end{eqnarray}  
where $I(y, \omega)$ denotes the spatio-spectral distribution of the harmonic intensity in Figs. \ref{fig4}(a) and (c). Similarly, the normalized pulse front tilt $\rho_{yt}$ is also evaluated from the spatio-temporal distributions $I(y, t)$ of the IAP in Figs. \ref{fig5}(a) and (c) according to
\begin{eqnarray}
\rho_{yt}=\frac{\iint{y t  I(y, t) dy dt}}{\sqrt{\iint{y^2 I(y, t)  dy dt}
		\iint{t^2 I(y, t)  dy dt}}}.
\end{eqnarray}  
The normalized spatial chirp $\rho_{y \omega}$ of the IAP generated in the near- and far-field in our two-color experiment are calculated to be -0.006 and 0.034, respectively. The normalized front tilts retrieved for the near- and far-field IAP are -0.023 and 0.014. Both the spatial chirp and front tilt of the generated IAP are very small, indicating a weak spatio-temporal distortion of the IAP in our experiment. The slight difference between the near- and far field results reveals a weak spatio-temporal coupling during the macroscopic propagation in our experiment. It can be expected that the spatio-temporal coupling will be more prominent for a highly-ionized dense gas medium, where the nonlinear propagation effect becomes more significant \cite{st1,st2}.

\section{Conclusion}

In conclusion, we have introduced an all-optical $in$ $situ$ method for the complete spatio-temporal characterization of IAPs. Our scheme relies on the weak perturbations induced by the perturbing laser on the HHG process in both time and space, which results in a spatial and also a frequency modulation in the far-field harmonic spectrum. From these modulations, the spatial and spectral phases of the harmonics can be accurately extracted by using a FROG-like method. With this method, the spatio-temporal
distributions of the IAP generated in a two-color driving field have been fully characterized in both the near- and far-field in experiment, from which the spatio-temporal coupling effect during the IAP generation is evaluated. This all-optical approach has a high efficiency in the data collection, thus can be applied to characterize the IAP generated in the experiment with a
low-repetition-rate laser source, where the acquisition of a photoelectron spectrum is usually time-consuming due to the space charge effects \cite{charge}. Moreover, our method is not restricted by the bandwidth of the harmonic spectrum. We have demonstrated in theory that the proposed method can be used to characterize broadband IAPs with  pulse durations less than 50 as. \\
\\
\textbf{Acknowledgments} 
\\
This work was supported
by the National Key Research and Development Program of China
(2019YFA0308300); National Natural Science Foundation of China
(91950202, 12074136, 11774109, 12021004); Fundamental Research Funds
for the Central Universities (2017KFXKJC002); Natural Science Foundation of Hubei Province (2021CFB330); and the Program for HUST Academic
Frontier Youth Team.\\
\\
\textbf{Data availability statement} 
\\
The data that support the findings of this study are available
upon reasonable request from the authors.

\end{document}